\documentclass{emulateapj}
\usepackage{natbib}
\usepackage{graphicx,graphics}

\makeatother

\shorttitle{The effect of bars and transient spirals}

\shortauthors{Saha et al.}

\begin{document}

\title{The effect of bars and transient spirals on the vertical heating in disk galaxies}

\author{Kanak Saha.$^{1,2}$, Yao Huan Tseng$^{1}$ and Ronald E. Taam$^{1,3}$  }
\affil{$^{1}$ Institute of Astronomy and Astrophysics, Academia Sinica - TIARA, Taiwan\\$^{2}$Max-Planck-Institut f\"ur extraterrestrische Physik, Garching, Germany \\ $^3$ Department of Physics \& Astronomy, Northwestern University, USA \\e-mail: kanak@asiaa.sinica.edu.tw}

\begin{abstract}

The nature of vertical heating of disk stars in the inner as well as the outer region 
of disk galaxies is studied. The galactic bar (which is the strongest non-axisymmetric pattern in 
the disk) is shown to be a potential source of vertical heating of the disk stars in the inner 
region. Using a nearly self-consistent high-resolution N-body simulation of disk galaxies, the 
growth rate of the bar potential is found to be positively correlated with the vertical heating 
exponent in the inner region of galaxies. We also characterize the vertical heating in the outer 
region where the disk dynamics is often dominated by the presence of transient spiral waves and 
mild bending waves. Our simulation results suggest that the non-axisymmetric structures are  
capable of producing the anisotropic heating of the disk stars.

\end{abstract}

\keywords{galaxies:evolution -- galaxies: kinematics and dynamics -- galaxies: structure -- galaxies: spiral -- stellar dynamics -- galaxies: haloes}

\section{Introduction}
The random velocities of disk stars in the solar neighborhood are known to increase with time since
the 1940's and this fact has been verified by recent Hipparcos observations. The
analysis of the Hipparcos data \citep{Binney2000} reveals that the stellar velocity dispersions
increase continuously (rather than episodically as claimed by \citet{Edvardsson1993}) with time
which is the well-known disk heating problem in our Galaxy. The phenomenon of disk heating is not
specific for our Galaxy since the heating of stellar disks is also known to persist in
external galaxies such as in NGC 2985, NGC 2460, and NGC 2775 \citep{Gerssen2000,Shapiro2003}. Although mechanisms have been identified for the disk heating, a full understanding has yet
to be achieved.  It is known that disk heating is anisotropic; that is, disk stars are preferentially heated along the plane in comparison to its perpendicular direction. Thus, the main problem in
disk heating is two-fold. The first is the requirement to explain the continual heating of disk
stars especially in the case of isolated galaxies for which tidal heating or external perturbations
can be ruled out.  Secondly, such a mechanism must be anisotropic in nature. Taken together, disk
heating remains a fundamental problem in galactic dynamics.

Historically, \citet{Spitzer1951,Spitzer1953} first showed that the secular increase in the
stellar velocity dispersions could arise as a result of the scattering of disk stars with giant
molecular clouds (GMCs) based on two dimensional calculations. Subsequently, there has been 
considerable development and progress in understanding the disk heating problem in galaxies. 
For example, \citet{Lacey1984} extended the previous theoretical work to a three dimensional disk and
found that the velocity dispersions increase with time, $t$, as $t^{\gamma}$ where $\gamma = 0.25$. 
In addition, it was demonstrated that transient stochastic spiral arms \citep{Barbanis1967,Carlberg1985} could also produce secular heating of the disk stars, however, such effects 
were  only effective in the plane of the galaxy. The effect of giant molecular clouds in combination 
with the effect of stochastic spirals could scatter the stars off the plane \citep{Jenkins1990} to 
account for both the radial and vertical heating. Alternatively, using an extensive 
orbit analysis of the 
stars subject to a 3D galactic bar potential, \citet{Pfenniger1984,Pfenniger1985} has shown that a large fraction of 
the phase space is chaotic and suggested that all the disk stars trapped in the chaotic phase space would 
be heated up in all directions.  Other possible candidate heating mechanisms include massive dark halo 
objects, e.g. black holes or dark clusters of mass $\sim 10^6$ M$_{\odot}$ \citep{Lacey1985, Carr1987} or a combination of giant molecular clouds and halo black holes \citep{Hanninen2002, Hanninen2004}. Within the framework of standard cosmological models, a spectrum of subhalos and their substructures could heat up the disk \citep{Sanchez1999, Ardi2003}. However, \citet{Font2001} found that 
the subhalos are not efficient perturbers for disk heating because the orbits of the subhalos rarely take 
them near the disk. Finally, satellite infall onto the galactic disk could produce an abrupt heating of the disk 
\citep{Quinn1993} or moderate heating could arise due to sinking satellites on prograde orbits \citep{Velazquez1999}. The many possible heating mechanisms of the disk stars has been 
reviewed by \citet{Pfenniger1993}. Note that while most mechanisms produce radial heating of the disk 
stars they have not been shown to provide a satisfactory explanation of the detailed nature of 
vertical heating.
In fact, studies focusing on the vertical heating are comparatively lacking. To be more precise, a generic 
and robust internal source remains to be identified which could be responsible for the continual vertical 
heating of the disk.

Here, we primarily concentrate on the issue of vertical disk heating. Galactic disks often consist of 
various non-axisymmetric patterns such as bars, spirals, warps, corrugations, and rings. These patterns 
may be growing or transient and, in general, may be time-dependent. It is natural to ask whether such 
patterns contribute to the disk vertical heating.  That is, what is the nature of vertical heating due,
for example, to a growing bar? How does a growing bar heat the disk stars in the vertical direction? 
Heating due to non-axisymmetric patterns are important for isolated galaxies as even for the Milky Way, 
there are prominent and characteristic signatures of non-axisymmetric patterns \citep{Dehnen1998} in the 
phase space in the solar neighbourhood.

In this paper, we explore possible mechanisms giving rise to continuous heating of the
disk stars in the vertical direction. In particular, we show that stellar bars that are formed nearly
spontaneously in disk galaxies are capable of efficiently scattering stars in the vertical direction.
Since transient spiral arms are often associated with bars, they also contribute to the disk
heating. In order to understand the nature of vertical heating of the disk stars, we have performed
a large number (statistically meaningful) of N-body simulations of model disk galaxies constructed
over a wide range of parameter space.

The paper is organized as follows. Section 2 describes the galaxy models used in the simulation. A 
description of the N-body simulation and the physical basis for the choice of our parameter space 
is presented in section 3.  The heating model used to interpret the numerical simulations is described in 
section 4. Section 5 describes the heating due to non-axisymmetric structures. The results on the correlation 
studies are presented in section 6. The discussion and conclusions are presented in section 7 and 8 respectively.

\section{Galaxy models}
The construction of a very thin equilibrium disk model involves a non-trivial procedure in N-body simulation.
However, using the self-consistent bulge-disk-halo model of \citet[hence KD95]{Kuijken1995} we are able
to construct extremely thin equilibrium model disks (with the ratio of scale-height to scale length $\sim
0.01$). Their prescription provides nearly exact solution of the collisionless Boltzmann and Poisson equations
which are suitable for studying disk stability related problems, allowing one to construct a wide range of
initial models from a large parameter space. All the components in our models are active (i.e.,
the gravitational potential of each component can respond to an external or internal perturbation) and, hence, provide a realistic representation
for the evolution and structure of the galaxies. Below we briefly describe each component of the model
separately for the sake of completeness.  For more details, the reader is referred to KD95.

A spherical live bulge is constructed from the King model \citep{Binney1987} and the corresponding distribution function (DF) is given by
\begin{equation}
  f_{b}(E)=\left\{
    \begin{array}{ll}
      \rho_{b}(2\pi\sigma_{b}^2)^{-3/2}
      e^{(\Psi_0-\Psi_{c})/\sigma_{b}^2} &\\
      \times \{e^{-(E-\Psi_{c})/\sigma_{b}^2}-1\}
        &\mbox{if} E<\Psi_{c},\\
      0   &\mbox{otherwise}.
    \end{array} \right.
\end{equation}

\noindent Here, the bulge is simulated using three parameters, namely the cut-off potential ($\Psi_c$), central
bulge density ($\rho_b$) and $\sigma_b$ governing the central bulge velocity dispersion.
The depth of the potential well is measured by $\Psi_0$. 

An axisymmetric live dark matter halo is simulated using the distribution function of a lowered Evans model
and is given as 
\begin{equation}
  f_{dm}(E,L_z^2)=\left\{
    \begin{array}{ll}
     [(A L_z^2 + B)e^{-E/\sigma_h^2} + C]\\
     \times (e^{-E/\sigma_h^2} -1) &\mbox{if} E<0,\\
      0   &\mbox{otherwise}.
    \end{array} \right.
\end{equation}

\noindent 
The velocity and density scales are given
by $\sigma_h$ and $\rho_1$ respectively. The halo core radius $R_c$ and the flattening parameter $q$ together with 
$\rho_1$ are contained in the parameters $A, B,$ and $C$. All the simulated halos are oblate in shape and kept at a 
constant value $q = 0.8$ for simplicity.

The disk distribution function is constructed using the approximate third integral given by $E_z = \frac{1}{2}
v_z^2 +\Psi(R,z) - \Psi(R,0)$, the energy of the vertical oscillations. This third integral is approximately
conserved for orbits near the disk mid-plane. The radial density of the disk is approximately exponential with
a truncation and the vertical density is chosen to depend exponentially on the vertical potential $\Psi_z(R,z)
= \Psi(R,z) - \Psi(R,0)$. The volume density of the axisymmetric disk is given by

\begin{equation}
\rho_d(R,z) = \frac{M_d}{8\pi h_z R_d^2} e^{-R/R_d} \mathop{\mathrm{erfc}}\left(\frac{R - R_{out}}{\sqrt{2}(R_{out} -R_{trun})}\right) f_d(z),
\end{equation}

\noindent where $f_d(z) = \exp(-0.8676 \Psi_z(R,z)/\Psi_z(R,h_z))$ governs the vertical structure of the disk, $\mathop{erfc}$ is the complementary error function. In the above equation, $M_d$ is the disk mass, $R_d$ is the scale length and $h_z$ is the scale height.

There are $13$ parameters required to construct a particular galaxy model. The galactic disk produced in this manner remains in equilibrium as long as the simulation runs. The dark matter halo generated using the lowered Evans 
model \citep{Evans1993} has a constant density core which is probably appropriate for the low surface brightness galaxies
which we simulate. Each model galaxy is constructed such that an initially almost flat rotation curve is
produced. The disk outer radius ($R_{out}$) is fixed at about $6.5 R_d$ and a truncation width $\sim 0.3 R_d$ is adopted
within which the disk density smoothly decreases to zero at the outer radius.

\section{N-body simulation}

The primary aim of the present study is to achieve an understanding of the nature and origin of vertical
heating of disk stars.  In order to gain the necessary insight, we perform a large number of simulations of isolated galaxies. 
The models are evolved for a sufficiently long period of time so as to examine the nature of vertical heating in different spatial regions of the disk.

Using the KD95 method, we build the initial conditions for all the model galaxies. The method is not fully
self-consistent because the disk distribution function is not known exactly. Each model galaxy consists of
a bulge, disk and dark matter halo as described above. Since the initial conditions are constructed based on
distribution functions, they are suitable for studying the long-term evolution of the non-axisymmetric patterns in
the disk. Building initial galaxy models with prescribed galaxy properties such as the Toomre $Q(r) =
{\sigma_r(r) \kappa(r)}/{3.36 G \Sigma(r)}$, ratio of dark-to-disk mass ($M_h/M_d$) or ratio of velocity
dispersions ($\sigma_z/\sigma_r$) is not straightforward because the bulge and the halo models are not
derived using their mass profiles but from their distribution functions. Hence, the relation between the
Toomre $Q$ parameter and $M_h/M_d$ or between $\sigma_z/\sigma_r$ and $M_h/M_d$ is not known from the outset.
Here, $\sigma_r, \sigma_z, \kappa,$ and $\Sigma$ denote the velocity dispersion in the radial and vertical
direction, epicyclic frequency, and surface density of stars at a given radius respectively.

The disk, bulge and halo parameters are chosen such that an almost flat rotation curve is produced in the outer
region. We assume a scale length for the disk to be $R_d = 4.0$ kpc. The radial and vertical forces are normalized
such that it produces the circular velocity $V_c = 200$ kms$^{-1}$ at about $2$ disk scale lengths. The unit 
of time is $2.1\times 10^7$ yr. The initial disk scale height ($h_z$) varies from $40$ pc to $400$ pc in our 
sample of model galaxies. The softening length for the disk particles is $10 $ pc, for the bulge particles 
$40$ pc and for the halo particles the softening length is $36$ pc. A total of $2.2 \times 10^6$ to $1.0 \times 
10^7$ particles has been used to simulate the model galaxies. The dark matter halo mass varies from 
$2$ to $40 \times M_d$ which gives, on average, a mass resolution of $\sim 10^5 M_{\odot}$. We have observed 
that with such a mass resolution the shot noise in the simulation reduces considerably and bending 
instabilities do not grow sufficiently to induce an outer disk warp during the evolution. Each model galaxy in our simulation has been evolved 
with the Gadget code \citep{Springel2001} which uses a variant of the leapfrog method for the time integration. The forces between 
the particles are calculated basically using the BH tree (with some modification) algorithm with a tolerance parameter 
$\theta_{tol} = 0.7$. The integration time step is $\sim 0.4$ Myr and each model galaxy is evolved for more than $5$ Gyr 
in our simulation. The orbital time scale at the disk half mass radius is $\sim 294$ Myr for the model mk97 (see Table.~\ref{paratab}). For convenience, we describe the parameter space for the simulations below.  

\subsection{Selection of the parameter space}
\begin{figure}
\includegraphics[angle=-90,scale=0.37]{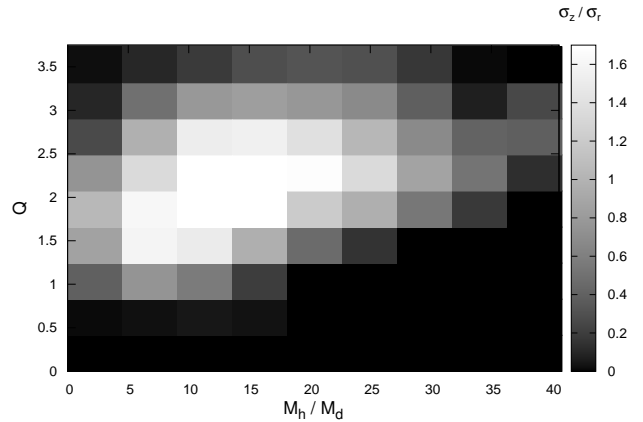}
\caption{The parameter space for all the models. Most of the galaxies in the sample are radially hot.
\label{fig:param}}
\end{figure}

\begin{figure}
\includegraphics[angle=270,scale=0.60]{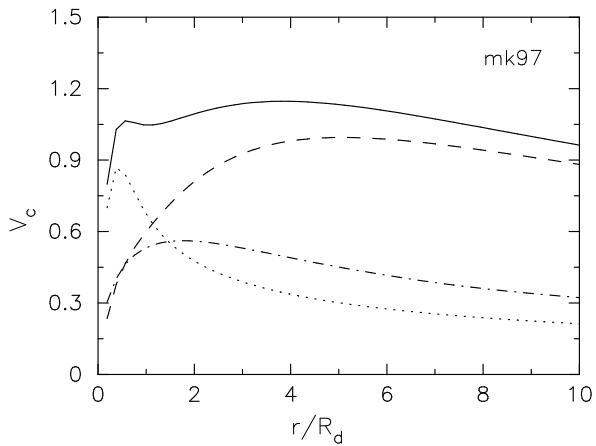}

\caption{Rotation curve for the model galaxy mk97. The solid line is the total rotation curve, dashed line 
is due to the dark matter halo, dash-dot-dash line for the disk and dotted line is for the bulge. }
\label{fig:rotc}
\end{figure}

A wide range of parameters in the 3-D space spanned by [$Q$,$\sigma_z/\sigma_r$,$M_h/M_d$] is considered. 
The exact relationship between these parameters is dependent on how an active dark matter halo interacts 
with the other active disk and bulge components.  However, we notice that if the central radial velocity 
dispersion is held constant, there appears to be a positive correlation between the Toomre $Q$ parameter 
and the dark halo-to-disk mass ratio ($M_h/M_d$) as expected from a simple understanding of the disk 
dynamics with a rigid dark halo. The initial thickness of the disks ($h_z/R_d$) in our sample of model 
galaxies ranges from thin ($h_z/R_d \sim 0.1$ or $0.15$) to superthin ($h_z/R_d < 0.1$). To give an example, 
the value of $h_z/R_d \sim 0.07$ in the superthin galaxy UGC 7321 \citep{matthews00}. We explore a wide 
region of the parameter space spanned by {$Q, M_h/M_d, \sigma_z/\sigma_r$} (see Fig.~\ref{fig:param}) and 
provide, here, the physical basis for our choice. Each model galaxy has a distinct $Q$ and $\sigma_z/\sigma_r$ 
radial profile; we quote the values of $Q$ and $\sigma_z/\sigma_r$ at the disk half mass radii (see Table~\ref{paratab} 
for representative models). Specifically, most of the galactic disks are vertically cold, but radially hot because we 
primarily aim to simulate low surface brightness (hence LSB) galaxies where the rotation curve is normally dominated by 
the dark matter halo \citep{deBlok2001}. In fact, many of the rotation curves in our sample (e.g., Fig.~\ref{fig:rotc}) are such 
that the dark matter dominates the disk rotation curve from the center of the galaxy. It is also known that 
the stellar disks of LSB galaxies are thinner than their high surface brightness (hence HSB) counterparts 
\citep{Bizyaev2004}. LSB galaxies are known to be poor in star formation activities despite the 
presence of neutral hydrogen gas as in otherwise normal galaxies. This may indicate that these galaxies 
are dynamically hot (with high $Q$ as most of the galaxies are in our sample) to prevent them from disk 
instabilities. Even if the disks suffer from instabilities, the resulting non-axisymmetric features could 
be young. In fact, the low metallicities, high gas-to-star mass ratios, and blue colors of most LSB galaxies 
indicate that these systems are probably younger than their HSB counterparts \citep{Vorobyov2009}.

\begin{table}
\caption[ ]{Initial conditions for the representative model galaxies (arranged in ascending order of $Q$) and the heating exponents computed at the end of each simulation.}
\begin{flushleft}
\begin{tabular}{ccccccc}  \hline 
Galaxy    & $Q$   & $\frac{\sigma_z}{\sigma_r}$  & $\frac{M_h}{M_d}$  &$\alpha_{in}$  & $\alpha_{out}$ &\\
models &              &     &   & &  &      \\
\hline
mk51A & 1.21 & 0.611 & 6.52 & 1.01 & 0.326 \\
mk56A & 1.45 & 0.594 & 10.15 & 2.59 & 0.557 \\
mk107 & 1.66 & 0.666 & 9.83 & 2.29 & 2.00 \\
mk104 & 1.77 & 1.480 & 15.80 & 1.67 & 0.67 \\
mk97 & 1.84 & 0.447 & 7.88 & 2.31 & 0.420 \\
mk15 & 2.06  & 0.237 & 5.34 & 1.13 & 0.809 \\
mk53 & 2.31 & 0.294 & 11.97 & 0.828 & 0.289 \\
mk52 & 2.33 & 0.618 & 14.47 & 1.81 & 0.707 \\
mk17 & 2.41 & 0.151 & 5.10 & 0.724 & 0.236\\
dmA & 2.37 & 0.433 & 12.97 & 0.776 & 0.588 \\
mk112 & 2.47 & 0.344 & 9.44 & 2.105 & 0.215 \\
mk5 & 2.50  & 0.149 & 9.85 & 0.406 & 0.448 \\ 
mk25  & 2.53  & 0.181 & 7.07 & 1.02 & 0.459 \\
mk63 & 2.64 & 1.160 & 14.12 & 0.865 & 0.902 \\
mk27 & 2.71 & 0.187 & 7.16 & 1.13 & 0.30 \\ 
mk64 & 3.00 & 1.110 & 16.50 & 0.723 & 1.39 \\
mk33 & 3.01 & 0.260 & 8.75 & 1.82 & 0.41 \\
mk43 & 3.05 & 0.606 & 11.52 & 1.30 & 1.63 \\
mk34 & 3.15 & 0.238 & 6.50 & 1.20 & 0.597 \\
mk58 & 3.16 & 0.850 & 16.75 & 1.27 & 1.02 \\ 
mk48  & 3.33  & 0.143 & 10.47 & 0.987 & 0.27 \\
mk55 & 3.73 & 0.540 & 27.60 & 1.16 & 1.10 \\
 
\hline
\end{tabular}
\end{flushleft}
\label{paratab}
\end{table}

\section{Heating Model}
We use the following simple expression (eq. 4) to fit all our simulation data, treating this functional form to 
quantify the effects of heating. It will be applied to the central region and to the outer region of the galaxy.

\begin{equation}
\sigma_z(r,t) = \sigma_0(r) + \sigma_1(r) \times (t/\tau_{orb})^{\alpha}.
\end{equation}

\noindent In the above equation, $\sigma_0$ and $\sigma_1$ are independent of time, and $\tau_{orb}$ is the
orbital time scale measured at the disk half mass radius. $\alpha$ is the heating exponent and is the
logarithmic time derivative of the vertical velocity dispersion $\alpha = d\ln{\sigma_z}/d\ln t$. We determine
the time evolution of $\sigma_z$ in the inner region ($< 1 R_d$) and in the outer regime ($\sim 4 R_d$) of the disk. The definitions for the two regions of the disk are maintained throughout the paper. The heating in each of these regions of the disk, characterized by apparently distinctive dynamical behaviors, is described by the resulting heating exponents given as $\alpha_{in}$ and $\alpha_{out}$ respectively. Higher values of $\alpha$ imply that the rate of heating is steeper as the galaxy evolves, whereas lower values indicate a gradual heating process. A thorough knowledge of these values are crucial for understanding the type of heating processes at work in the galaxy and possibly for identifying the mechanism responsible for the heating. The fitting is performed using a 
robust non-linear least squares fitting algorithm due to Levenberg and Marquardt \citep{Markwardt2009, More1978}.
The best fit model parameters are chosen based on the minimum value of the chi-square ($\chi_{min}^2$). Using the resulting parameters, an useful estimate for the time scale of heating can be obtained from the above formula and 
this can also be compared with the relaxation time scale for the model under consideration.

Let us consider that in time $\tau_h$, the change in the vertical velocity dispersion $\Delta \sigma_z \sim \sigma_0(r)$ at a particular radius in the disk. Then differentiating eq.[4], it can be shown that 

\begin{equation}
\tau_h = \left[{\frac{\sigma_0(r)}{\alpha  \sigma_1(r)}}\right]^{1/\alpha} \tau_{orb}.
\end{equation}

We evaluate this time scale for different models presented here. In comparison, the energy relaxation time 
scale for the N-body model with a total number of particles $N$, size $R_s$ and softening $\epsilon_s$ can 
be written following \citet{Huang1993} as
\begin{equation}
\tau_{relax} = \frac{N \times \frac{R_s}{R_{1/2}}}{32\pi \log(R_s/\epsilon_s)} \tau_{orb},
\end{equation}
where we have used the crossing time scale $\tau_{cross} = \frac{R_s}{2\pi R_{1/2}} \tau_{orb}$, assuming that 
the circular velocity $V_c$ remains almost flat beyond $R_{1/2}$; $R_{1/2}$ denotes the half-mass radius. 
Considering $N = 5 \times 10^6$, $R_s = 10$, $R_{1/2} =2.5$ and softening parameter $\epsilon_s =0.003$, 
we have $\tau_{relax} = 5.6\times 10^4 \tau_{orb}$. 

Previously, disk heating has been modelled as a diffusion process in the velocity space. If the diffusion 
coefficient $D_z$ remains constant in time, it can be shown that the vertical velocity dispersion evolves 
\citep{Lacey1991} as

\begin{equation}
\sigma_z = (\sigma_{z0}^{1/\gamma} + D_z \frac{t}{\tau_{orb}})^{\gamma}. 
\end{equation}

Note that the above formula is applicable to the radial dispersion also. The exponent $\gamma = 1/2$ denotes 
heating due to the halo black holes \citep{Lacey1985}. In addition to eq. [4], we also apply the 
diffusion model to fit our simulation data to some models for comparison and evaluate the corresponding diffusion coefficient $D_z$. Let 
us consider that $\tau_d$ is the time scale over which the change in the vertical velocity dispersion  
$\Delta \sigma_z \sim \sigma_{z0}$. Then $\tau_d$ can be obtained as follows

\begin{equation}
\tau_d =  (\sigma_{z0}^2/D_z)\tau_{orb}.
\end{equation}

We denote $\tau_d$ as the diffusion time scale. In general, $\tau_d$ and $\tau_{relax}$ are comparable. Internal 
evolution of a disk galaxy is driven by various non-axisymmetric perturbations such as bars or spirals which 
redistribute the energy and angular momentum within the disk and/or between the halo and the disk. The time 
scale for this process is longer than the dynamical time scale ($\tau_{dyn}$) and is usually known as the 
secular evolution time scale ($\tau_{sec}$) \citep{Kormendy2004}. $\tau_{sec}$ could be a few to 
few $10$s of orbital time scales ($\tau_{orb}$). In ordering these time scales, $\tau_{dyn} < \tau_{orb} < 
\tau_{sec} < \tau_{d} \lesssim \tau_{relax}$. We will use these definitions and compare the relevant time 
scales in the following to determine the relation between $\tau_h$ and the other time scales.

\begin{figure}
\newlength{\figwidth}
\setlength{\figwidth}{\textwidth}
\addtolength{\figwidth}{-\columnsep}
\setlength{\figwidth}{0.5\figwidth}
  
  \begin{minipage}[t]{\figwidth}
    \mbox{}
    \hskip 98pt
    \vskip 6pt
    \centerline{\includegraphics[width=0.72\linewidth,angle=0]{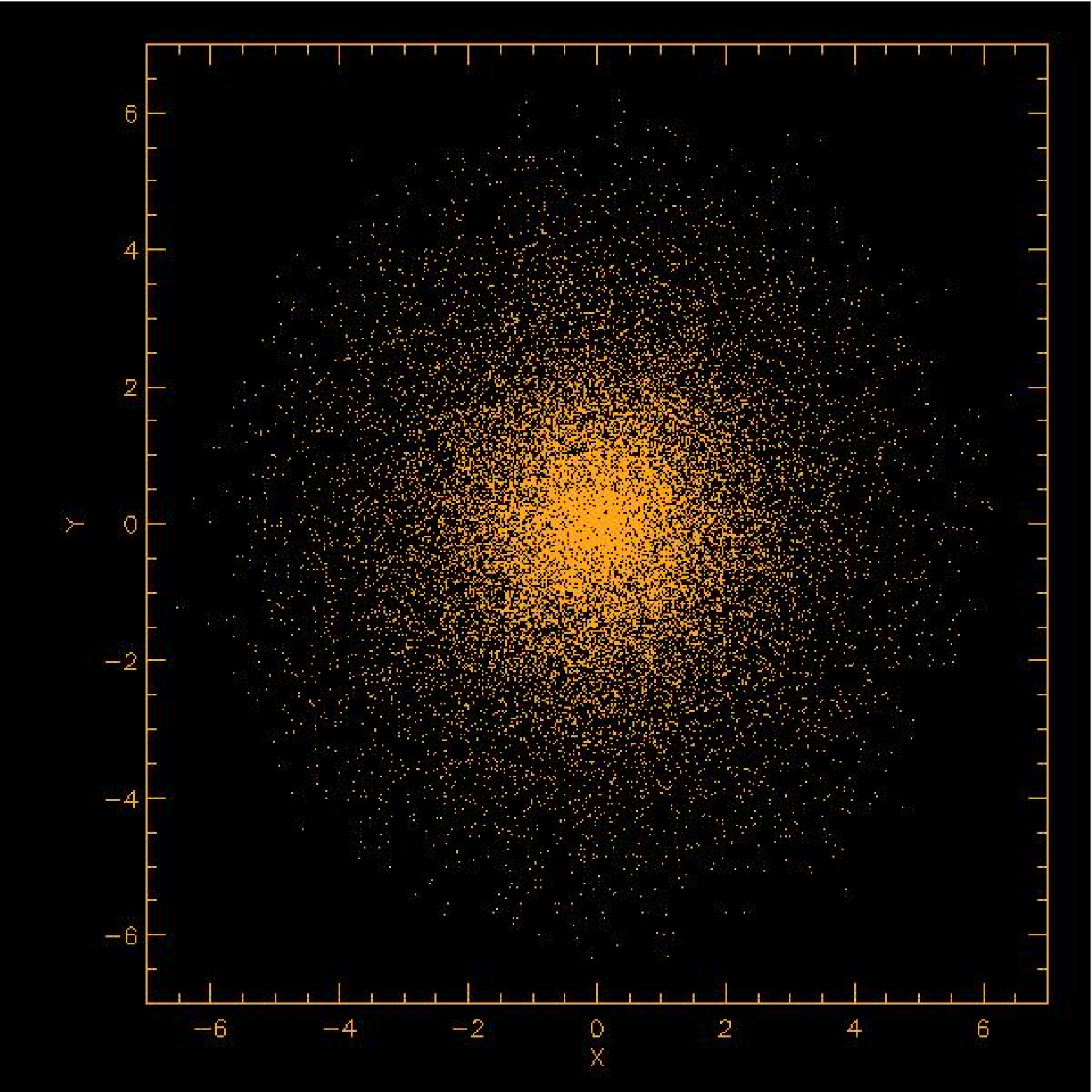}}
  \end{minipage}
  \hfill
  \begin{minipage}[t]{\figwidth}
    \mbox{}
    \vskip -178.98pt
    \hskip -115pt
    \centerline{\includegraphics[width=0.262\linewidth,angle=0]{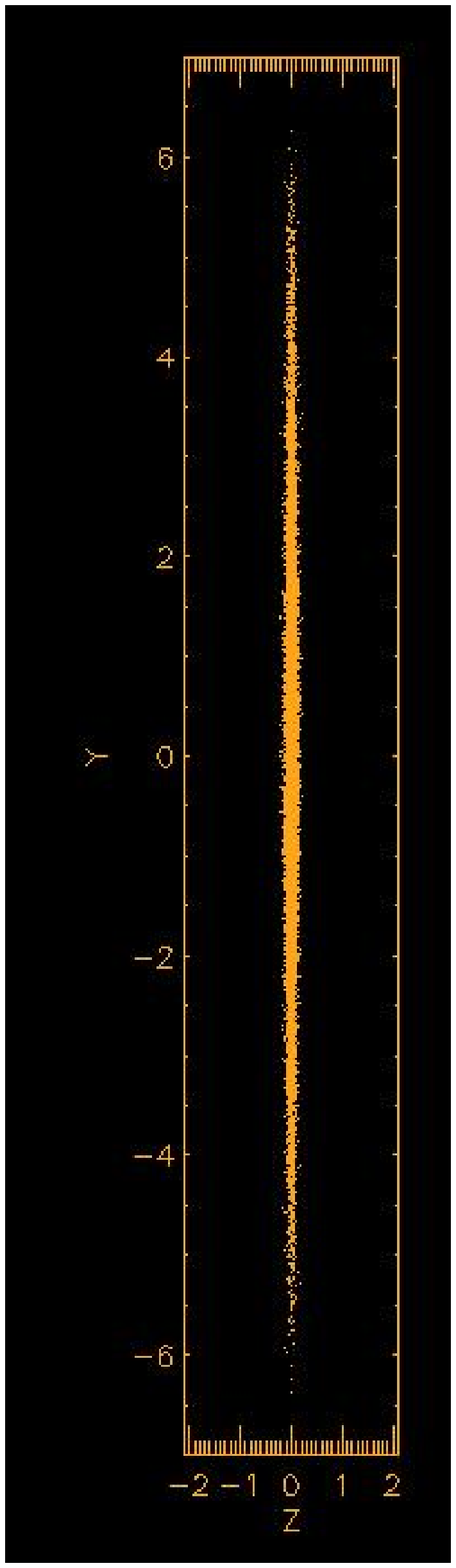}}

  \end{minipage}

  \begin{minipage}[t]{\figwidth}
    \mbox{}
    \vskip -61.0pt
    \hskip -0pt
    \centerline{\includegraphics[width=0.72\linewidth,angle=0]{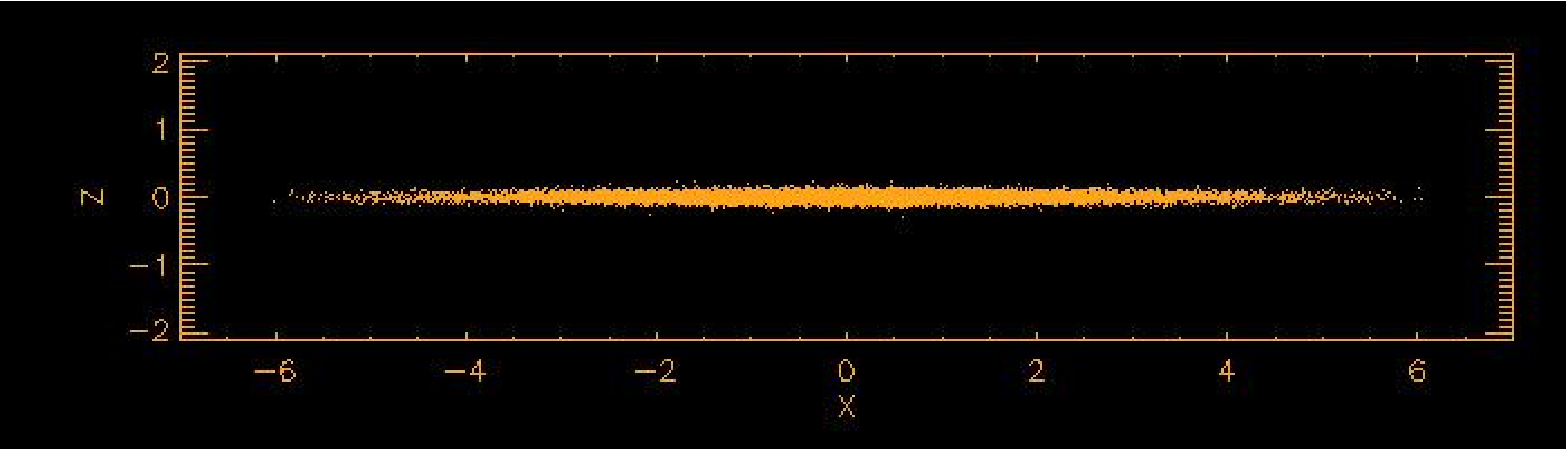}}

  \end{minipage}

\caption{ Particle image of mk97 showing the face-on and the two edge-on views of the initially axisymmetric disk (at T=0.)}

\label{fig:mk97t0}

\end{figure}

\section{Non-axisymmetric structures and disk heating}
It is well established that stars undergo continuous heating in both the radial and vertical directions in
galaxies as they age. Although there has been substantial progress in understanding the radial heating,
there is little progress for the vertical heating. Spiral structures have played a significant role in the
studies of disk heating in the radial direction.  We note, however, that a steadily rotating quasi-stationary spiral pattern heats the disk stars only negligibly \citep{Binney1987}. In this case, sufficient 
heating of disk stars requires the spiral structures to be stochastic in nature. Transient, swing-amplified spiral density waves \citep{Fuchs2001} or multiple spiral density waves \citep{Minchev2006} have recently 
been shown to be effective in the radial heating of disk stars.

The primary mechanisms suggested for the vertical heating of the disk stars in early works range from 
a combination of stochastic spirals and giant molecular clouds \citep{Jenkins1990} to resonant 
heating of the trapped stars in the chaotic phase space in the presence of a 3D bar potential 
\citep{Pfenniger1985}. More recently, other mechanisms have been suggested including a bending instability of the entire disk or of the bar \citep{Sotnikova2003} and repeated impact of globular clusters \citep{Vandeputte2009}, although it produces only a small change in the vertical velocity dispersion. Disk galaxies are rich in non-axisymmetric structures, which may form either by means of mergers, tidal interaction or internal instabilities. 
Examples include bars, transient spirals, rings \citep[e.g. NGC 1433, NGC 6300][]{Buta2001}, lopsidedness \citep{Saha2007}, bending waves manifested in the form of a warp \citep[e.g. ESO121-G6, NGC 4565][]{Saha2009} or corrugation \citep[e.g. in IC 2233][in NGC 5907]{Matthews2008,Saha2009} of the disk mid-plane. These non-axisymmetric instabilities often drive the disk away from equilibrium and are generally time dependent. Each of these features are, in principle, a potential source of heating of the disk stars. Since 
our understanding of the interactions of these features with the surrounding live dark matter halo is not 
well understood, it is difficult to disentangle the signatures of each of these non-axisymmetric modes on 
the net heating of a group of stars.  However, some progress has been made especially for our Galaxy in 
which it has been shown based on the analysis of the Hipparcos data that there exists a gap due to 
the $2:1$ resonance with the bar in the Galactic UV plane \citep{Dehnen2000}.

In the following, we study the nature of the vertical heating of the disk stars due to bar growth and 
transient spirals.  We also briefly discuss the effect of halo noise and the corrugation and/or 
bending waves on the net heating process. We carry out a series of N-body simulations to first 
systematically identify the sources of disk heating in our model galaxies which are free from tidal 
interaction. About 70 simulations have been performed covering a wide range of the above mentioned 
simulation space and a statistical attempt to characterize and understand the nature of vertical heating 
in various regions of the disk has been carried out. 

\begin{figure}
\setlength{\figwidth}{\textwidth}
\addtolength{\figwidth}{-\columnsep}
\setlength{\figwidth}{0.5\figwidth}
  
  \begin{minipage}[t]{\figwidth}
    \mbox{}
    \hskip 98pt
    \vskip 6pt
    \centerline{\includegraphics[width=0.72\linewidth,angle=0]{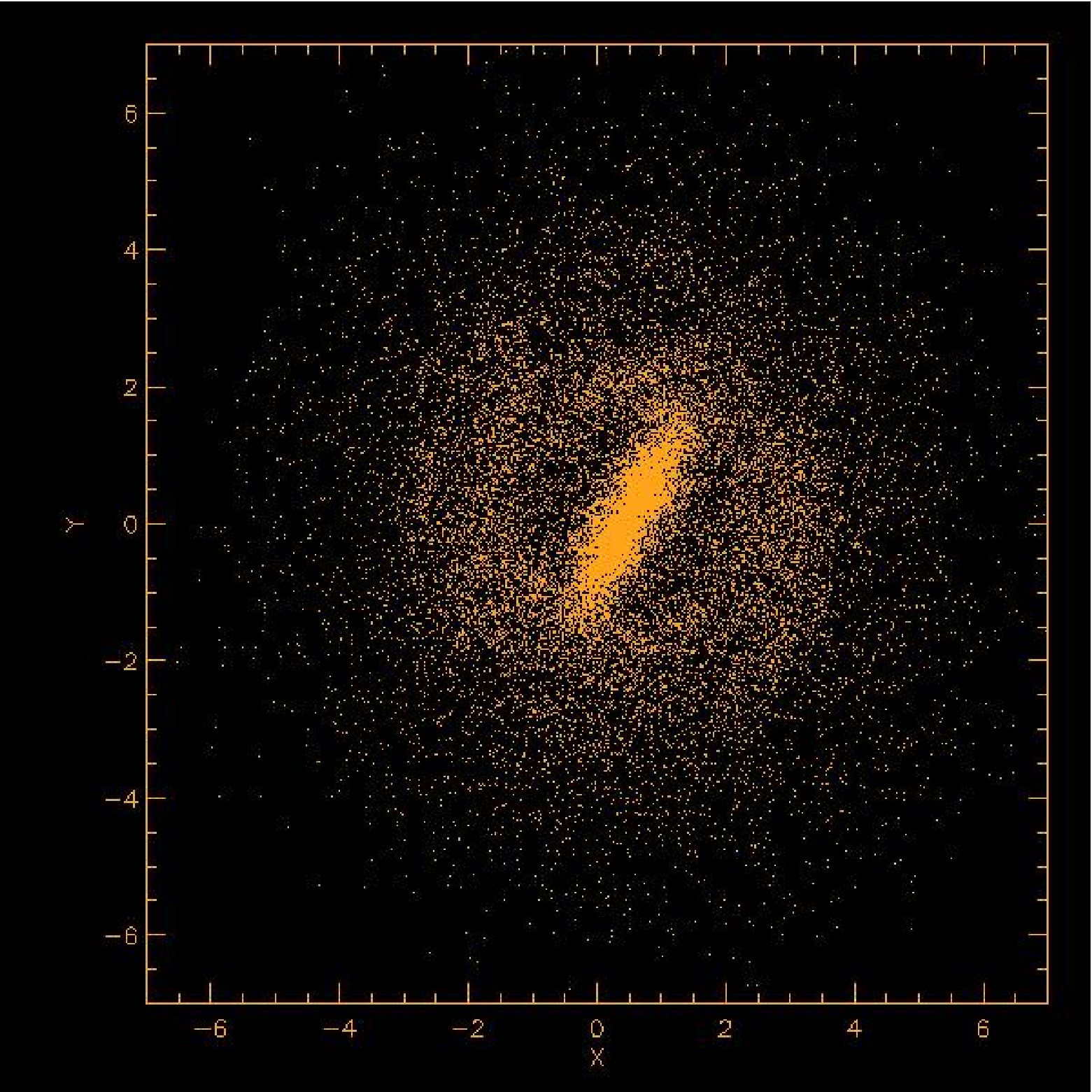}}
  \end{minipage}
  \hfill
  \begin{minipage}[t]{\figwidth}
    \mbox{}
    \vskip -178.98pt
    \hskip -115pt
    \centerline{\includegraphics[width=0.262\linewidth,angle=0]{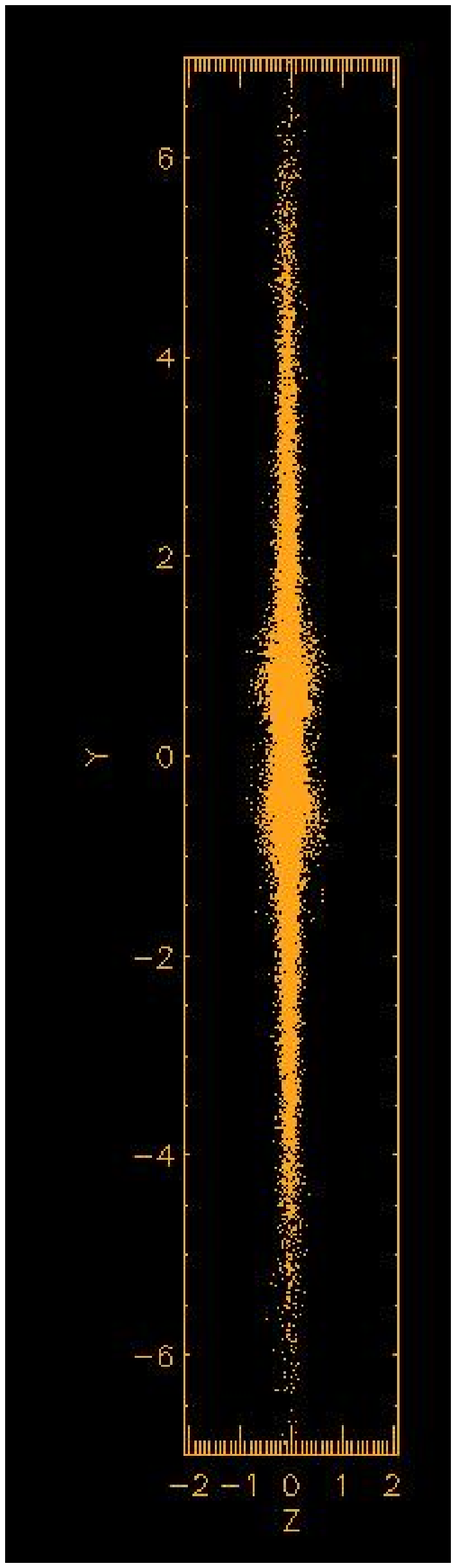}}

  \end{minipage}

  \begin{minipage}[t]{\figwidth}
    \mbox{}
    \vskip -61.0pt
    \hskip -0pt
    \centerline{\includegraphics[width=0.72\linewidth,angle=0]{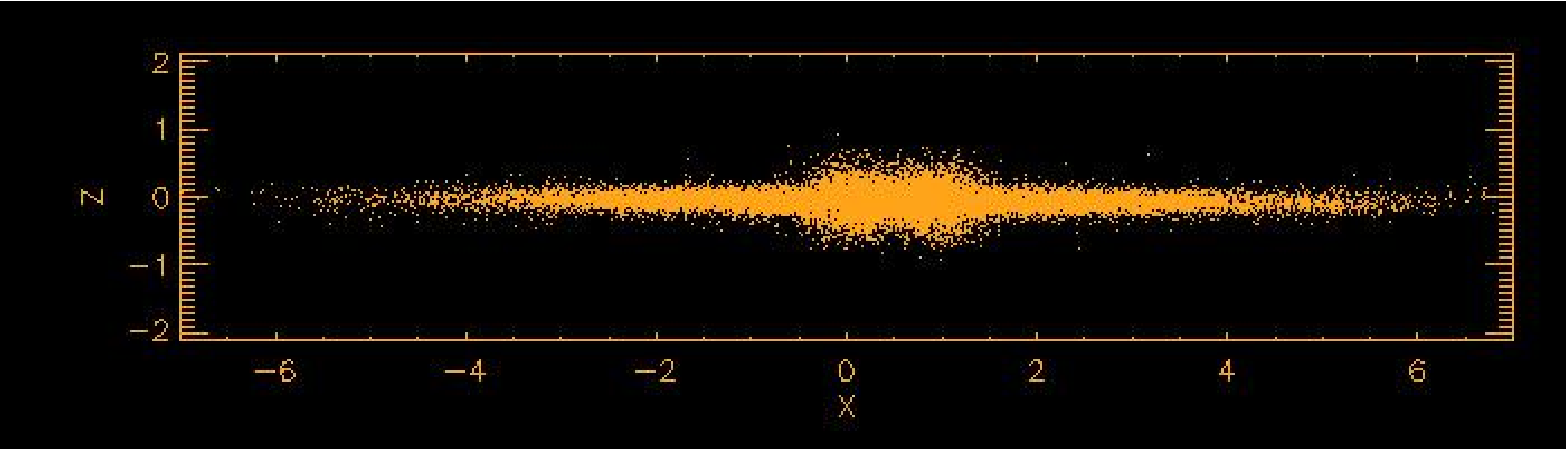}}

  \end{minipage}

\caption{ Particle image of mk97 at the end of the simulation (T=5.4 Gyr). The disk forms a peanut bulge which lasts till the end.}
\label{fig:mk97tf}
\end{figure}

\begin{figure}
\includegraphics[angle=270,scale=0.60]{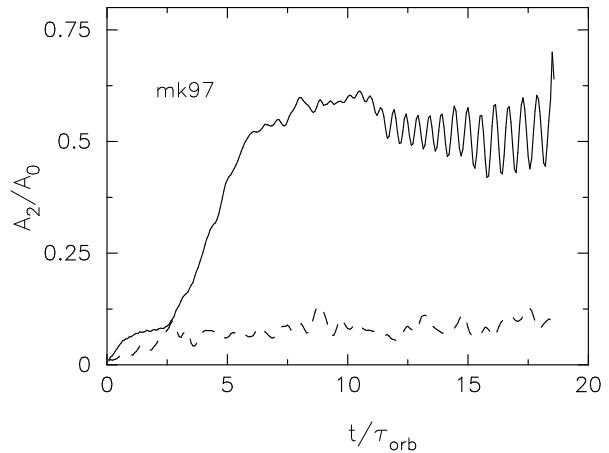}
\caption{The amplitude of the bar is shown as a function of time.  The solid line represents a {\it type-I} bar 
generated in the center of the model mk97. The dashed line indicates the amplitude of spirals in the outer 
region.} 
\label{fig:bsmk97}
\end{figure}

\begin{figure}
\includegraphics[angle=270,scale=0.60]{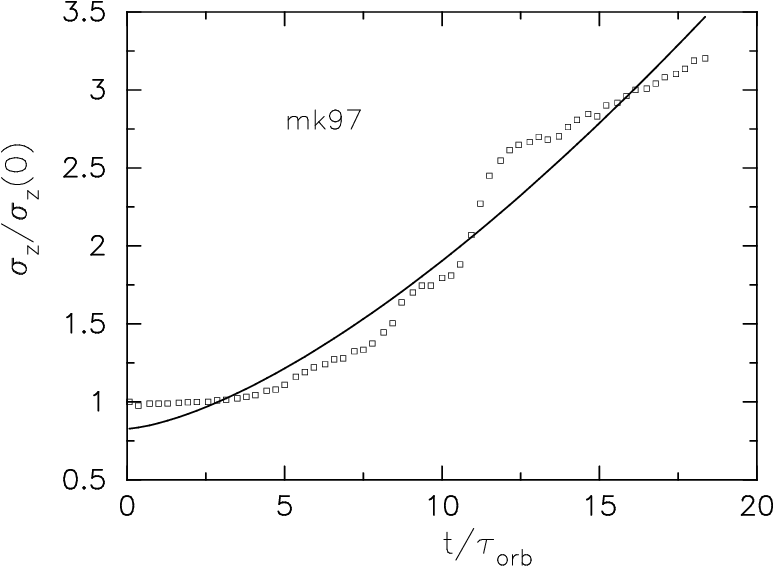}
\caption{The vertical heating in the inner region ($< 1 R_d$) of the galaxy model mk97 as a function of time. The solid 
line is the model fitted to the simulation data. $\sigma_z(0)$ is the initial value of the vertical dispersion 
at this radius.}
\label{fig:sigzinmk97}
\end{figure}

\begin{figure}
\includegraphics[angle=270,scale=0.60]{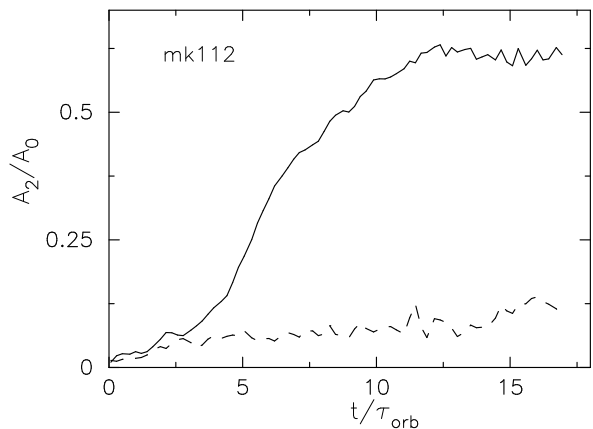}
\caption{Same as Figure~\ref{fig:bsmk97}  except for model mk112.}
\label{fig:bsmk112}
\end{figure}

\begin{figure}
\includegraphics[angle=270,scale=0.60]{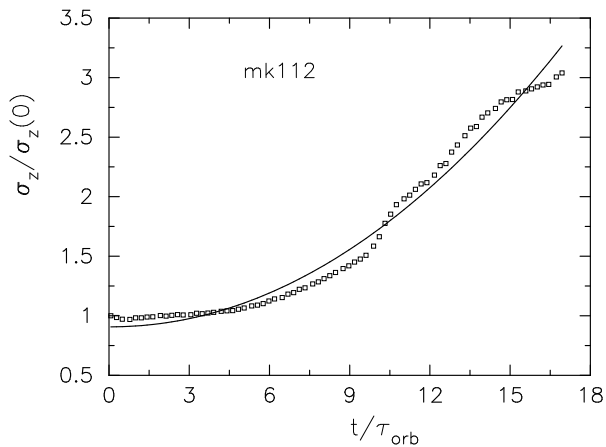}
\caption{Same as Figure~\ref{fig:sigzinmk97} except for model mk112.}
\label{fig:sigzinmk112}
\end{figure}

\subsection{Vertical heating due to bar growth}
Although most of the galaxies in our sample are radially hot, they form a bar at the center during its time evolution. 
Some of the model galaxies form a bar within about $3$ to $4$ rotation time scales, while for others it takes more 
than about $10$ rotation time scales to reach nearly the same amplitude. It is found that a bar forms even for 
galaxies characterized by $Q \sim 3.7$ (model mk55) in our simulation, implying that it is difficult for a rotating disk 
within a live dark matter halo to avoid forming bars at the center. Hence, bar formation appears to be quite common and 
generic in disk galaxies and a live dark matter halo plays a significant role. As shown by \citet{Athanassoula2002} 
a live dark matter halo supports, rather than suppresses, the bar instability in the disk. In contrast, bar 
formation in the disk could be suppressed in the presence of a non-responsive rigid dark matter halo \citep{Hohl1976}. The growth of the bar depends on the initial condition and grows via non-linear processes as the disk stars interact with the 
live dark matter halo particles. The nature of this interaction is yet to be completely understood although 
recent N-body simulations by \citet{Dubinski2009}, \citet{Sellwood2009}, and \citet{Klypin2009} have revealed insights on the nature of bar dynamics in disk galaxies.  For our choice of parameters and initial conditions, we do not find any unique trends for the bar growth in our simulation. The sample of galaxies in our N-body simulations differs from typical galaxy models studied previously (where $Q$ is normally held constant throughout the disk or $Q \le 2.0$ in general). Overall, we observe that the growth of bars in 
our sample falls under two broad distinct categories. In the first category, the bar grows quickly to a peak amplitude within $\sim 5$ rotation time scales which is about an order of magnitude less than $\tau_{sec}$ and nearly reaches saturation or starts growing again (for convenience, we call these {\it{type-I bar}}). In the second category, the bar continues to grow but slowly as compared to {\it type-I} and does not show any tendency to saturate (we call these {\it{type-II bar}}). {\it Type-II} bars grow on a secular evolution time scale to reach the same amplitude as in {\it type-I}. These two main features of bar growth are illustrated in the appropriate figures below, although there are some intermediate cases present in our simulation. In many cases, the bar undergoes the well known buckling instability \citep[and references therein]{Combes1981, Pfenniger1991, Raha1991, Debattista2006, Martinez2006} following which the bar takes the form of a peanut and/or X shaped bulge. The buckling instability driven by 
the anisotropy in the velocity dispersions is an important phase of the bar evolution as it leads to the formation of a 
pseudo-bulge \citep{Kormendy2004} directly influencing the secular evolution 
of disk galaxies. However, the condition for the onset of this instability and the exact underlying cause of 
this phase remains a matter of further investigation. For example, recent work by \citet{Martinez2004} has shown that the buckling instability weakens the bar and the resulting peanut shaped phase lasts for several Gyrs in their simulation. We also find that the peanut shaped phase of the bar is long lasting, typically surviving till the end of our simulation ($> 5$ Gyr) producing boxy/peanut/X-shaped structure in the central region. The morphological evolution of the disk in model mk97 is depicted through 
Fig.~\ref{fig:mk97t0} and Fig.~\ref{fig:mk97tf}. These snapshots are taken at time $T=0$ and $T=5.4$ Gyr which denote the beginning and the end of the simulation respectively. The initially axisymmetric disk evolves to form a bar which eventually transforms into a peanut morphology. Inspection of the evolution of the bar-amplitude reveals a diverse behavior while it is in the peanut phase; in some cases the amplitude continually grows and 
in others, it saturates. While the bar strength continues to increase, the evolution of its pattern speed is not always 
correlated in the sense that the pattern speed does not always decrease. In particular, the bar pattern speed in some 
cases remains nearly unchanged with time while its amplitude grows (as in mk17, mk33, mk104). A similar 
anti-correlation between the bar growth and pattern speed evolution has been discussed by \citet{Valenzuela2003} and more recently by \citet{Villa2009}. However, there are also the 'normal' cases where the pattern speed decreases with time, while the bar grows to a higher amplitude, by losing angular momentum to the halo via dynamical friction. We notice that the bar growth is slow (typically a factor of $5$ to $10$ less) in cases where the pattern speed is observed to remain nearly constant in time. Overall, the emerging diverse behaviour of the bar characteristics precludes detailed understanding for inferring gross physical properties 
(e.g. bar size, velocity dispersions) of the disk based on a few simulation results. As a consequence, the 
results of a large number of simulations have been used to approach the problem in a statistical sense.      

\begin{figure}
\includegraphics[angle=270,scale=0.60]{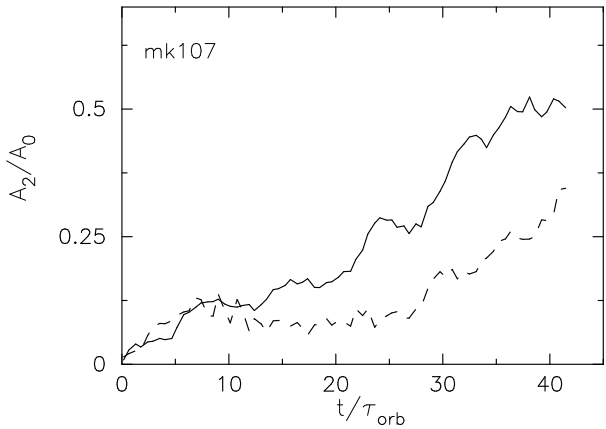}
\caption{ Solid line represents the amplitude of the bar ({\it type-II}) generated in the center of the model mk107, while 
the dashed line is for the amplitude of the spiral in the outer region.}
\label{fig:bsmk107}
\end{figure}

\begin{figure}
\includegraphics[angle=270,scale=0.60]{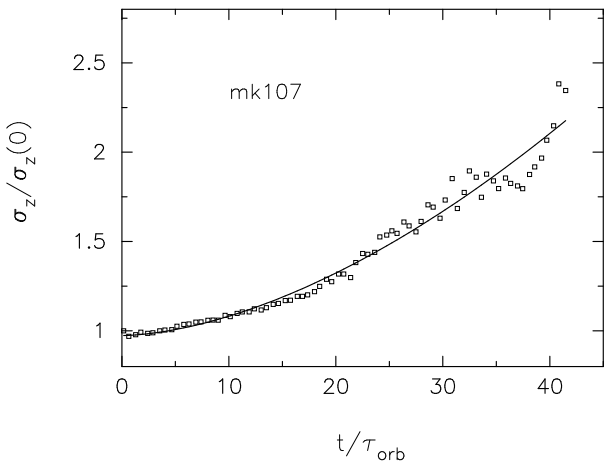}
\caption{Vertical heating in the central region of the galaxy model mk107. Solid line is the model fitted to 
the simulation data. $\sigma_z(0)$ is the initial value of the vertical dispersion at this radius.}
\label{fig:sigzinmk107}
\end{figure}

For each of the model galaxies, we compute the $m=2$ Fourier component ($A_2$) of the particle distribution 
from each N-body snapshot using the following formula:
\begin{equation}
A_{m}(r,t) = \frac{1}{N_r}\sum_{j=1}^{N_r}e^{i m \phi_j(r,t)} 
\end{equation}
In the above equation $\phi_j(r,t)$ is the phase of the $j^{th}$ particle at position $r$ and time $t$ and 
$N_r$ represents the number of particles used. Since $A_m$ is a complex number, the modulus of the 
$m^{th}$ Fourier component is obtained by $|{A_m}| = \sqrt{\Re{(A_m)}^2 + \Im{(A_m)}^2}$ \citep{Sellwood2009} and the corresponding phase is given by $\phi_m = \frac{1}{m}\tan^{-1}{\Im{(A_m)}/\Re{(A_m)}}$, ($m \ne 0$). 
In general, the radial variation of $A_2$ shows a pronounced peak corresponding to a bar in the 
central region of the disk and often a second peak indicating the presence of a spiral perturbation in the 
outer parts of the disk. The variation of the peak amplitude of $A_2$ in the inner region with time reveals 
the growth of a bar in the galaxy. The amplitude $A_2$ is normalized to the amplitude of the axisymmetric 
mode ($m=0$ Fourier component). In all the subsequent figures, we present the smoothed normalized $A_2$. 
For a wide range of model bars, the following linear regression model is fit to the time evolution:
\begin{equation}
\log A_2(r,t) = A_2^0 + \beta \log(t/\tau_{orb}),
\end{equation}
where the slope $\beta$ is the logarithmic time derivative of the bar amplitude. The linear model in the 
log-log plot translates to the bar amplitude evolving as $A_2(r,t) = A_2^0(t/\tau_{orb})^{\beta}$.  In the 
following, $\beta_b$ represents the growth rate of a bar and $\beta_s$ represents the growth rate of a 
spiral (in the outer region).

The time evolution of the normalized bar amplitude in model mk97 is illustrated in Fig.~\ref{fig:bsmk97}. The 
bar amplitude nearly saturates through a series of high frequency oscillations while undergoing the buckling 
phase. Such oscillations have also been reported in previous simulations by \citet{Valenzuela2003} and \citet{Villa2009}.  
Our estimate shows that the typical period of such oscillations is $\sim 1/2 \times 
\tau_{orb}$ or less than $\tau_{orb}$ in general. This roughly implies that the amplitude oscillates with 
a frequency close to $\kappa$, the radial epicyclic frequency.  The vertical velocity dispersion ($\sigma_z$) 
of the stars in the inner region of the disk is plotted as a function of time in Fig.~\ref{fig:sigzinmk97} for 
model mk97. From Fig.~\ref{fig:sigzinmk97}, it can be seen that $\sigma_z$ changes its slope at $\sim$ 12 
rotation time scales, nearly following the overall amplitude evolution of the bar. The  best fit parameters 
of the heating model (eq.[4]) are $\sigma_0 =0.09$, $\sigma_1 = 1.7 \times 10^{-4}$ and $\alpha_{in} = 2.31$. 
Using these parameters in eq.(5), we obtain $\tau_h = 10.56\times \tau_{orb} \ll \tau_{relax}$. The heating 
time scale due to the bar growth is found to lie in the range $\tau_{dyn} < \tau_h \ll \tau_{relax}$. 
We have also applied the diffusion model given by eq.(7) to the 
inner region of mk97 and find that the heating model given by eq.(4) provides a better fit than the diffusion 
model. In particular, the diffusion coefficient $D_z = 5.9 \times 10^{-4}$ and $\gamma = 109$ are found to be 
absurdly large. The resulting time scale $\tau_d = 1.5 \times 10^3 \tau_{orb}$.    

Interestingly, the temporal variation of $\sigma_z$ does not reflect any such oscillation while the radial 
velocity dispersion undergoes oscillation with nearly the same frequency as the bar amplitude beyond about 
$10$ rotation time scales. 

The growth of a bar in model mk112 is similar to that in model mk97, although the disk in model mk112 is hotter 
($Q = 2.47$) than mk97. Fig.~\ref{fig:bsmk112} shows the normalized bar amplitude and the time evolution of $\sigma_z$ in the disk inner region is shown in Fig.~\ref{fig:sigzinmk112}. 
The best-fit parameters from model mk112 are found to be similar to mk97.

In the case of model mk107, the bar does not reach saturation and continues to grow as shown in Fig.~ \ref{fig:bsmk107}. 
The vertical velocity dispersion in the inner region of the disk grows more gradually (Fig.~\ref{fig:sigzinmk107}) 
following the evolution of the bar. Note that the overall increase in the vertical velocity dispersion is less than 
that of model mk97 and mk112. We find that the heating model given by eq.(4) better approximates the simulation data 
than the diffusion model. The resulting diffusion coefficient $D_z = 2.2 \times 10^{-4}$ is similar to model mk97 and 
$\tau_d = 4.4 \times 10^3 \tau_{orb}$. For models mk97, mk107 and mk112, we find that, in general, the diffusion model 
overestimates the dispersion values in the inner region of the disk. 

In all our simulations, the vertical velocity dispersion increases following the growth of the bar. 
Note that all the three galaxy models (e.g. mk97, mk107 and mk112) undergo a phase of buckling instability during 
their evolution, but {\it the disk stars are continually vertically heated prior to and during/after the buckling instability 
phase.} This fact suggests that the vertical heating of the disk stars in our simulations is not strictly related 
to the bar buckling phase (e.g. the disk in model mk17 does not go through buckling instability within the simulation 
time period but still shows continuous heating in the vertical direction). However, we note that the rate of vertical 
heating during the bar buckling phase is generally higher than in phases without such buckling. 

In the presence of a rotating bar ($m=2$) potential, the planar motion of the disk stars can be coupled with the vertical oscillation (parametric resonance) at the location of the 
vertical resonances $\nu/(\Omega(r) - \Omega_b) = m/n$, where $\nu$ is the vertical 
frequency, $\Omega_b$ is the bar pattern speed, $n$ is the number of vertical oscillations. For $m=2$ perturbation, the $n = \pm 1$ represent the vertical Lindblad resonances. For the $n=\pm 2$ resonances, the retrograde orbits in the inner region of the galaxy couple efficiently with the vertical motion through the 'Binney instability strips' \citep{Binney1981}. \citet{Pfenniger1985} has shown that indeed these $2:1$ resonances inside the bar play a significant role in trapping stars, leading to rapid diffusion in the vertical direction. In the presence of a growing bar potential, the disk stars are subjected to a complex instability and the main effect of such an instability is to promote fast diffusion of the low angular momentum, high z amplitude and energetic stars to the nearby halo. The instability is shown to grow as the bar perturbation grows. Basically, the presence of a growing bar potential (i.e. a time dependent potential in the galaxy) breaks the time invariance symmetry (because bar growth is an irreversible proprocess) and hence the jacobi integral which in turn enhances the ergodicity considerably allowing the stars to visit most of the chaotic phase space \citep{Pfenniger1984, Pfenniger1985}. This provides a basic understanding of the heating mechanism in the presence of a growing bar. Of course, in N-body simulation the nature of bar growth is quite diverse as mentioned in the beginning of this section. The resonant heating of the disk stars in the presence of a rotating (with changing pattern speed) growing (in amplitude) bar potential is  probably more complex, leading to shifting locations of the resonances in the inner part of the galaxy and at the same time promising in the context of vertical heating in the disk. It is natural to examine how the vertical heating of the disk stars is correlated with the growth of bar, the nature of the vertical heating due to a growing bar, and the typical exponent for the vertical heating.  In section 6, we seek the possible correlations between the bar growth rate ($\beta_b$) and the inner heating exponent ($\alpha_{in}$). 
 
\begin{figure}
\includegraphics[angle=270,scale=0.60]{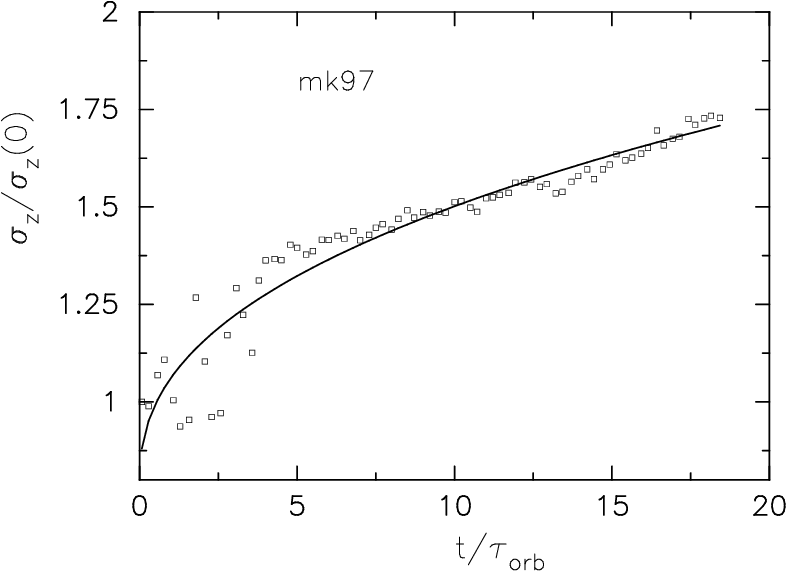}
\caption{The vertical heating in the outer region ($\sim 4 R_d$) of the galaxy model mk97 as a function of time. The solid 
line is the same model fitted to the simulation data. $\sigma_z(0)$ is the initial value of the vertical dispersion at 
this radius.}
\label{fig:sigzoutmk97}
\end{figure}

\begin{figure}
\includegraphics[angle=270,scale=0.60]{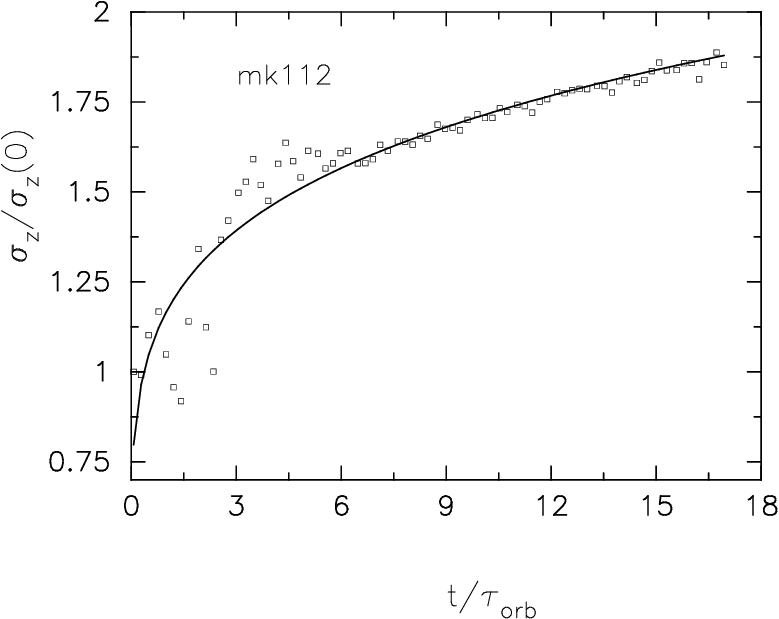}
\caption{Same as figure~\ref{fig:sigzoutmk97} except for galaxy model mk112.}
\label{fig:sigzoutmk112}
\end{figure}

\begin{figure}
\includegraphics[angle=270,scale=0.60]{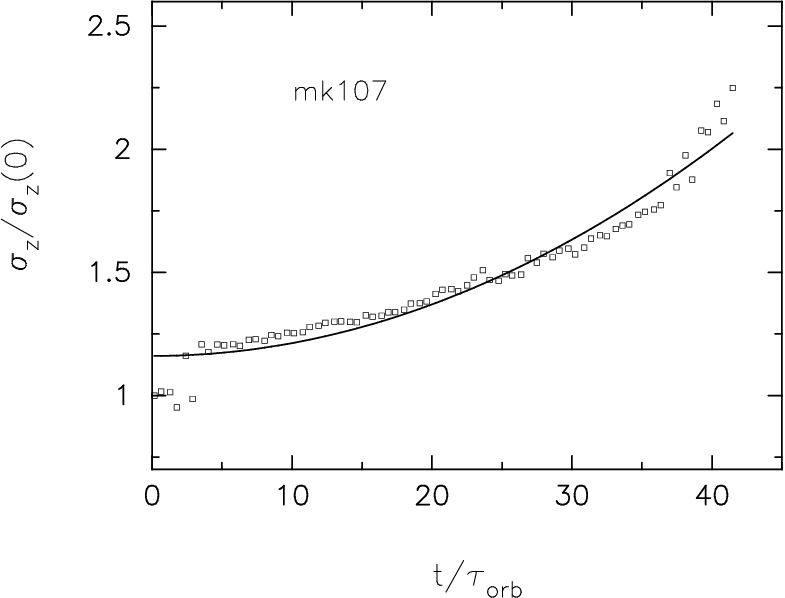}
\caption{Vertical heating in the outer region ($\sim 4 R_d$) of the galaxy model mk107. Solid line is the same 
model fitted to the simulation data. $\sigma_z(0)$ is the initial value of the vertical dispersion at this radius.}
\label{fig:sigzoutmk107}
\end{figure}

\subsection{Outer disk heating and transient spirals}
Strong two-armed spirals are difficult to excite and form in a radially hot disk galaxies. Using the Fourier 
decomposition of the disk surface density, however, many transient spirals (often with no steady pattern speed) are 
detected in our model galaxies. These transient spirals are often diffuse and generally weak in radially hot disks 
(Fig.~\ref{fig:bsmk97} and Fig.~\ref{fig:bsmk112}) as compared to the relatively strong spiral formed in the 
radially cold disk model mk107 (see Fig.~\ref{fig:bsmk107}). These spirals are primarily confined to the outer region of the stellar disk, and we examine whether they play a role in the vertical heating of the disk stars in this region. In Figs.~\ref{fig:sigzoutmk97}, \ref{fig:sigzoutmk112} and \ref{fig:sigzoutmk107}, we present the time evolution of the vertical velocity dispersion in the outer region of the disk for models mk97, mk112 and mk107 respectively. The slope of the heating curve in 
model mk107 (concave) is quite different from the other two models (convex) presented here. In the later stages of the 
evolution, $\sigma_z$ in model mk107 increases faster than that in mk97 and mk112 (where $\sigma_z$ nearly saturates). 
Using the best-fit parameters of our heating model fitted to the outer region of mk97, we find that $\tau_h = 100 \tau_{orb}$ 
where $\sigma_0 = 0.025$, $\sigma_1 = 8.6\times 10^{-3}$ and $\alpha = 0.42$. Applying the diffusion model in the outer region of mk97, we obtain a diffusion coefficient $D_z = 6.35\times 10^{-8}$ and $\gamma = 0.214$ and the corresponding time scale 
$\tau_{d} = 1.3 \times 10^4 \tau_{orb}$, which is similar to the relaxation time scale ($\tau_{relax}$) quoted in section 4. 
In the case of model mk107, the best-fit parameters from our heating model yields $\tau_h = 33.7 \tau_{orb}$. The diffusion model 
in this case gives $D_z = 8.48\times 10^{-4}$ and $\tau_{d} = 1.15\times 10^{3} \tau_{orb}$. These time scales indicate 
that the process of vertical heating in the outer region of the disk is slower in comparison to the inner region. We find that in 
the presence of a relatively strong spiral in the outer region, disk stars are heated to a greater degree in the vertical 
direction and the heating rate is found to be faster. The actual heating time scale, $\tau_h$, (in Gyr) in mk107 is lower by a 
factor of $6$ in comparison to models mk97 or mk112. Although, the exact mechanism through which stars are vertically heated in 
the outer region is not clear, growing spirals definitely play a significant role.

The outer region of the stellar disk is not just simply described by the transient spirals, as corrugation waves or mild 
warps of the disk midplane are often present. It is known that the transient spirals can heat the disk efficiently 
in the radial direction, but poorly in the vertical direction. However, GMCs could redistribute the random energy of the 
stars in the vertical as well as in the radial direction through scattering processes. In the absence of GMCs or massive 
objects (such as halo black holes or dark clusters) in the outer parts of the galaxy, one seeks a process whereby the 
heating due to spirals or some other feature in the disk is redistributed in the vertical direction. Possible candidates, 
for example, include corrugation waves, large scale warps of the stellar disk or, in general, vertical motion of the disk 
stars coupled with the transient spirals. The growth and development of these bending waves or rather warps is known 
to a large extent to depend on the noise in the N-body system. We have verified that as the number of particles is 
increased (say by a factor of $10$) in the system, the growth of these bending waves is significantly reduced because of 
the substantial decrement in the Poisson noise arising from particle discreteness. With an average mass resolution of 
$\la 10^5 M_{\odot}$, only weak bending waves or mild warping of the stellar disk are present; comparatively strong 
warps can arise with a poorer mass resolution $\ga 10^6 M_{\odot}$. Thus, it is unlikely that the bending waves alone 
can heat the disk stars in the outer region with higher mass resolution. However, the possibility exists that the non-linear 
coupling between the weak transient spirals and the weak bending waves \citep{Masset1997} could redistribute energy in the 
vertical direction.  As mentioned earlier, an interesting behaviour about the bars in radially hot disks inside a live dark 
matter halo was noticed.  It was found that the {\it type-I} and {\it type-II} bars trigger transient spirals with distinctive 
characteristics in the outer disk. In the case of a {\it type-II} bar, the transient spiral continues to grow in amplitude, and 
this process continues as the bar grows (see Fig.~\ref{fig:bsmk107}). This is found to be the case even in one of our hottest disk models (mk55 for which $Q = 3.73$, see Table~\ref{paratab}). Because of the growing transient spirals, we find that the radial heating dominates over the vertical one throughout the disk.  On the other hand, the amplitude of the spirals nearly saturates in the case of models mk97 and mk112 (see Fig.~\ref{fig:bsmk97} and Fig.~\ref{fig:bsmk112}) where the bar is of {\it type-I}. Note that in this case, the radial heating nearly saturates throughout the disk (see Fig.~\ref{fig:heatcomp_rzin} and Fig.~\ref{fig:heatcomp_rzout}). Overall, there is a parallel development prevailing in the inner and the outer regions of the disk whereby a bar and spiral grow respectively.  
 
\subsection{Anisotropic heating}
An important aspect associated with the heating of disk stars is its anisotropic nature as can be inferred from the 
fact that the observed ratio of the vertical to radial velocity dispersion i.e. $\sigma_z/\sigma_r < 1$  \citep[as in NGC 488][]{Gerssen1997}, as found to occur for the solar neighbourhood in the Galaxy. This fact raises many questions.  For 
example, why do disk galaxies have $\sigma_z/\sigma_r < 1$? Can the heating mechanism preserve an initial 
$\sigma_z/\sigma_r$ value?  What causes the stars to be heated anisotropically? 

\begin{figure}
\includegraphics[angle=270,scale=0.60]{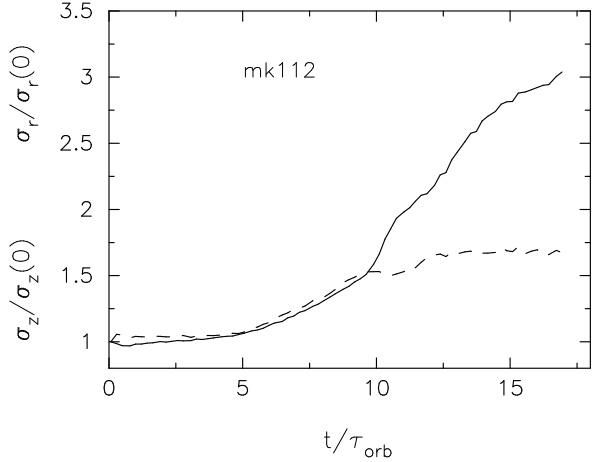}
\caption{ Comparison of radial and vertical heating in the inner region in model mk112. $\sigma_z(0)$ and $\sigma_r(0)$ are the values of initial dispersions at that radius. Solid line denotes vertical heating and dashed line the radial heating.}
\label{fig:heatcomp_rzin}
\end{figure}

\begin{figure}
\includegraphics[angle=270,scale=0.60]{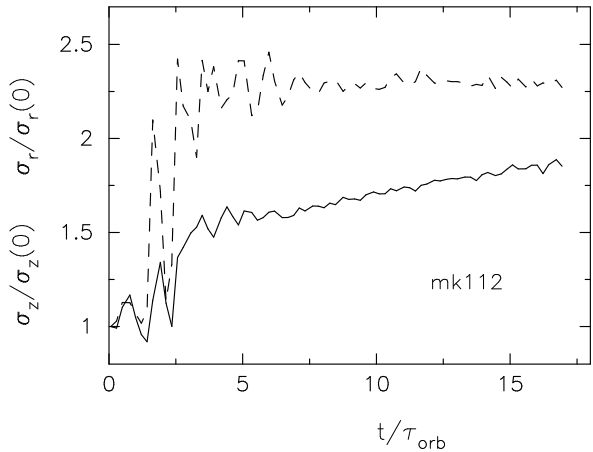}
\caption{ Comparison of radial and vertical heating in the outer region of model mk112. $\sigma_z(0)$ and $\sigma_r(0)$ are the values of initial dispersions at that radius. Solid line denotes vertical and dashed line the radial heating.}
\label{fig:heatcomp_rzout}
\end{figure}

To gain some insight into these questions, we illustrate in Fig.~\ref{fig:heatcomp_rzin} and Fig.~\ref{fig:heatcomp_rzout} 
the evolution of the radial and vertical velocity dispersions in the inner and outer region of the model galaxy mk112. 
It can be seen that the overall nature and the rate of heating is different for the radial and the vertical directions. In 
the inner region of mk112, the vertical heating is more effective than the radial heating and in the outer regime the radial 
heating saturates while the vertical heating continues slowly. We find this trend to be the case for most of the 
galaxy models which are radially hot and in which the bar evolves to saturation. On the other hand, for model mk107 
which is relatively cold ($Q = 1.66$), the radial heating dominates over the vertical heating throughout the disk due to 
the presence of a relatively strong spiral perturbation. It is clear that the non-axisymmetric structures (such as bars 
or spirals) are essential for disk heating.

\begin{figure}
\includegraphics[angle=270,scale=0.60]{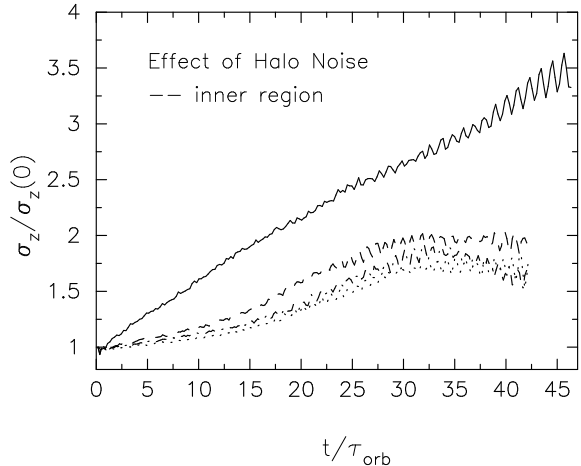}
\caption{Vertical heating curves in the central region of the galaxies with different number of dark matter 
halo particle leading to different halo mass resolution. Solid line is for 0.5M, dashed line for 2M, dash-dot line for 3M, and dotted line for 5.0M particles}
\label{fig:hnoise_in}
\end{figure}

\subsection{Halo noise}
\begin{figure}
\includegraphics[angle=270,scale=0.60]{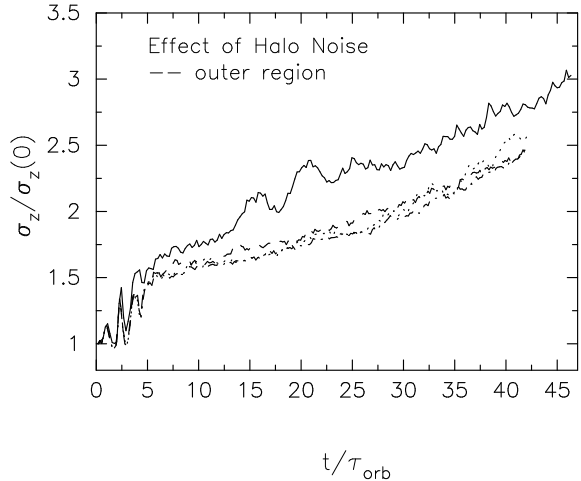}
\caption{Vertical heating curves in the outer region ($\sim 4 R_d$) of the galaxy models with different number 
of dark matter halo particles. Solid line is for 0.5M, dashed line for 2M, dash-dot line for 3M, and dotted line for 5.0M particles}
\label{fig:hnoise_out}
\end{figure}

The role of a live or responsive dark matter halo has been emphasized by \citet{Athanassoula2002} in the context of bar 
formation in N-body simulations. Unlike the case of a rigid dark matter halo, a live halo facilitates the growth of 
the bar in an N-body disk, likely through resonant interactions with the disk particles. On the other hand, the finite 
number of particles employed in the N-body simulation can be an important source of disk heating as well.  Because of 
the finite number of particles, two-body relaxation processes \citep{Velazquez2005} can be effective in heating the disk 
stars efficiently. It is highly desirable to eliminate or, at least, minimize the Poisson fluctuation in the halo 
particle distribution by increasing the number of halo particles employed in the simulation as permitted by the 
available computational resources. To quantify this effect, we performed four additional simulations on the galaxy model 
dmA (see Table~\ref{paratab}) with different number of halo particles ($N_h$). There is an order of magnitude variation in the number of halo particles between models with the poorest resolution (0.5 million (M) particles) and the highest resolution. The intermediate two simulations use $2$ and $3$ million halo particles. In Fig.~\ref{fig:hnoise_in} and Fig.~\ref{fig:hnoise_out}, we show the four vertical heating curves in the inner region and in the outer region respectively. 
It is found that the vertical heating curves nearly converge for models with $N_h > \mathrm{few} \times 10^6$. The convergence on the number of particles in our N-body simulation is in accordance with the conclusions reached in the recent work by \citet{Dubinski2009} in the context of bar formation and evolution. Comparison of Fig.~\ref{fig:hnoise_in} and Fig.~\ref{fig:hnoise_out} reveals that the effect of halo noise is more important in the central region than in the outer region. In order to quantify this trend, the time scales $\tau_h$ in the inner and the outer region were determined. We find that an order of magnitude increase in the halo particle number increases $\tau_h$ by at least a factor of $10$ in the inner region and $\sim 2$ in the outer region. In contrast, we find different behaviour when the temporal variation of the radial velocity dispersion in the disk was examined. Because these models are dominated by spiral arms in the outer region and we find that by increasing the number of halo particles (from 0.5 M to 5 M) the spiral arms became even stronger. This resulted in a lower radial dispersion in models with lower number of halo particles compared to the models with higher number of halo particles. In the case of model mk112, the spiral arms are rather weak and the radial heating of the disk stars could have resulted from the halo noise. However, we find that the radial dispersion remains nearly constant over more than $10$ rotation time scales (see Fig.~\ref{fig:heatcomp_rzin} and Fig.~\ref{fig:heatcomp_rzout}). Based on these facts, it is possible to conclude that the noise due to the halo (with $N_h >$ few million particles) is neither effective nor the dominant source of vertical heating in the outer region of the disk. 

In summary, we find a positive correlation between the growth of the bar at the galactic center and the vertical heating 
taking place in the central few kpc region. The formation of a bar is almost unavoidable in a stellar disk embedded in a 
live dark matter halo under the wide range of physical parameter space that has been explored. Bar formation occurs even 
in a very hot galaxy with its time for onset delayed at higher numerical resolution. In addition, analysis based on the 
four simulations on model dmA reveals that the strength of the spiral arms also increases with higher mass resolution. It 
has been noted that the vertical heating exponent is generally larger in the outer regions of a galaxy in the presence of 
strong spirals.  Hence, the transient spirals in the 3D disk which are common in our simulations not only play a 
significant role in radial heating, but also the vertical heating of disk stars. 

\begin{figure}
\includegraphics[angle=270,scale=0.60]{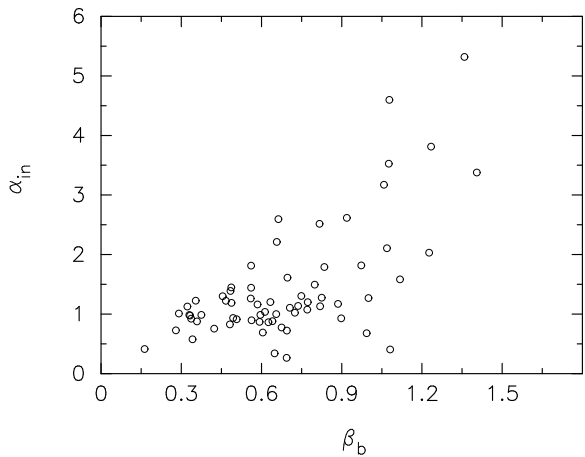}
\caption{Vertical heating exponent ($\alpha_{in}$) in the inner region ($\sim 1 R_d$) for all the models. Dependence on the bar growth rate ($\beta_{b}$).}
\label{fig:corr_barsigz}
\end{figure}

\section{Correlations}
In the absence of a clear understanding of the physical processes responsible for the vertical heating of disk stars, 
we approach the problem by seeking possible correlations between the average physical properties (e.g., bar strength 
versus vertical heating rate) from our fairly large sample of N-body galaxies. To obtain a measure of the strength and slope 
of possible relations, we carry out a statistical correlation, measuring it between two variables x and y by the Pearson's 
product-moment correlation coefficient defined as follows:

\begin{equation}
\mathop{\mathrm{Corr}}(x,y) = \frac{\mathop{\mathrm{cov}}(x,y)}{\delta_x \delta_y} = \sum_{i}^{N_s}{\frac{(x_i - \bar{x})(y_i - \bar{y})}{(N_s -1)\delta_x \delta_y}}
\end{equation}

In the above equation, $\mathop{cov}$ represents the covariance and $\delta$ denotes the standard deviation of the 
particular variable in the subscript. The summation is carried over the sample size $N_s$. According to the 
Cauchy-Schwarz inequality the correlation can not exceed 1. If the value of the correlation is positive, it 
indicates an increasing linear relationship and in the case of negative value, it denotes a decreasing linear 
relationship. 

In Fig.~\ref{fig:sigzinmk97}, Fig.~\ref{fig:sigzinmk112}, and Fig.~\ref{fig:sigzinmk107} the time evolution of the vertical 
velocity dispersion in the inner region of the disk is illustrated. In comparison with the evolution of the bar, it 
appears that the vertical velocity dispersion of the stars roughly follows the time evolution of the bar. Thus, for 
individual galaxies, there is a trend for the vertical heating, and it is interlinked with the growth of the 
bar. However, the exact relationship between the two and the reason for the vertical heating of the disk stars closely 
following the growth of the bar is unknown. In the absence of a definitive answer, we seek to determine whether the trend found in individual galaxies is generic. For example, does it depend on the initial condition, and is this trend robust when examined on a large sample of galaxies with different initial conditions? We note that these are difficult questions to answer because galaxies with varying initial conditions evolve quite differently. As pointed out recently by \citet{Sellwood2009} a live dark matter halo interacts with the disk in a more complicated way than previously envisioned.
  
In Fig.~\ref{fig:corr_barsigz}, the dependence of the vertical heating exponent ($\alpha_{in}$) in the inner region of 
the galaxy on the growth rate of the bar ($\beta_b$) is illustrated for our sample of model N-body galaxies. The figure 
clearly shows that the heating exponent in the inner region of the disk is strongly positively correlated with growth 
rate of the bar with a value of $\mathop{Corr}(\alpha_{in},\beta_b) = 0.647$. As pointed out earlier, when 
examined on a case by case basis, greater vertical heating results in galaxies which host stronger bars (e.g., models 
mk56A, mk52, mk112). Our analysis suggests that as the strength of the bar increases, the disk stars continue to be 
heated vertically, irrespective of the bar's pattern speed evolution. The heating exponents tend to be higher in 
models with comparatively less massive dark matter halos. However, for more massive halos, the vertical heating 
exponent ($\alpha_{in}$) tends close to unity. On the other hand, in the absence of a strong bar, the heating exponent 
is lower, $\alpha_{in} \sim 0.5$.

\begin{figure}
\includegraphics[angle=270,scale=0.60]{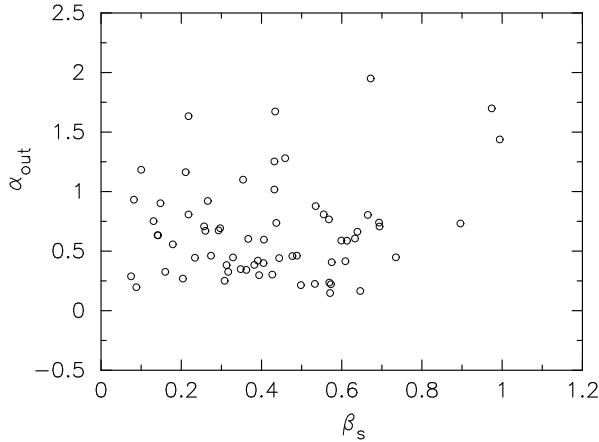}
\caption{Vertical heating exponent ($\alpha_{out}$) in the outer region ($\sim 4 R_d$) for all the models. Dependence on the growth rate of spirals ($\beta_{s}$).}
\label{fig:corr_spiralsigz}
\end{figure}

A weak correlation is found between the growth of spirals and the heating exponent in the outer region as can be seen 
from Fig.~\ref{fig:corr_spiralsigz} corresponding to a value of $\mathop{Corr}(\alpha_{out},\beta_s) = 0.177$. Its low value 
indicates that there may be more than one mechanism operating. Although the correlation between the heating exponents 
and the growth exponents for the spirals is not strong, higher exponents are associated with the presence of stronger 
spirals (e.g., model mk43, mk55, mk64) and the overall vertical heating is low when the spiral arms are 
more diffuse and weak (e.g., models mk5, mk51A, mk48). The vertical heating in the outer region is less extensive than the 
central region. The heating exponents in the outer region in many models are very different from $\alpha_{out} = 0.5$ 
indicating that the disk stars are not heated purely by massive objects in the halo or by the transient spirals, however, 
the exact source(s) which are responsible for the vertical heating in the outer region have yet to be identified.

\begin{figure}
\includegraphics[angle=270,scale=0.60]{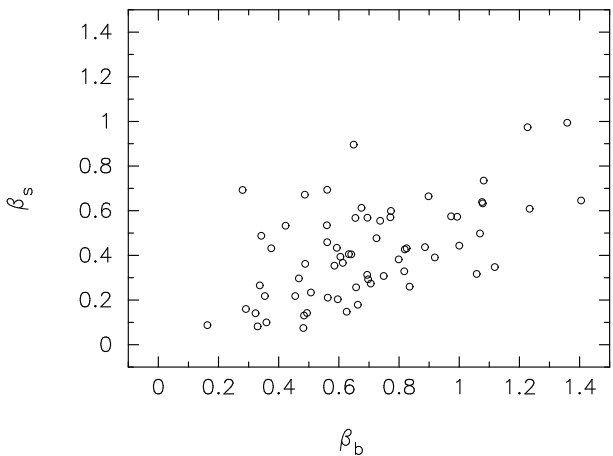}
\caption{Correlation between the growth rates of bars ($\beta_{b}$) and spirals ($\beta_{s}$) for all the model galaxies. The figure shows that atleast weak spirals are formed in the disk whenever there is a bar.}
\label{fig:corr_barspiral}
\end{figure}

On the other hand, we find that there is a relatively strong correlation (see Fig.~\ref{fig:corr_barspiral}) between the 
strength of the bars and the strength of the spirals in the disk with a value of $\mathop{Corr}(\beta_b,\beta_s) = 0.565$. This correlation 
indicates that in the presence of a bar there is always at least a weak spiral present in the outer disk. This may also suggest that such 
weak transient spirals are triggered by the presence of a bar in the disk central region. However, the vertical heating in 
the inner region and the outer region is not well correlated. As the correlation between the two is very weak 
$\mathop{Corr}(\alpha_{in},\alpha_{out}) =0.10$, the source of vertical heating in the inner and the outer region of the 
disk is likely to differ.  

\section{Discussion}
We have studied the nature of vertical heating of the disk stars in a large sample of radially hot galaxies as a function 
of halo mass and $\sigma_z/\sigma_r$ spanning a wide range. For an individual galaxy, we find that the stars in the 
inner region are heated vertically as long as a bar continues to grow. Such correlation is quite generic and found 
to be robust when examined on a larger sample of model galaxies constructed from different initial conditions. Bars 
form in galaxies with diverse initial conditions but do not evolve in a comparable fashion likely reflecting their 
non-linear interaction with the surrounding dark matter distribution. However, we find that on a broad category 
there are two kinds of bar: {\it type-I} (growing rapidly and approaching saturation) and {\it type-II} (growing slowly with no 
evidence of saturation) as mentioned in section 5.1. Heating due to the growth of a bar differs from that due to a 
diffusion process since it occurs on a time scale of a few rotations in the disk central region. The diffusion time 
scale in a stellar system is similar to the relaxation time scale because diffusion mainly causes the stars velocity 
to change via two body encounters. Thus from our analysis, it can be said that the diffusion approximation is likely to 
be inappropriate for the vertical heating of disk stars in the central region. 

\subsection{Growth of boxy/peanut bulge}
The boxy/peanut bulges are commonly believed to originate from the galactic bar via the vertical heating \citep[and references therein]{Combes1990, Pfenniger1990, Pfenniger1993, Athanassoula2005, Athanassoula2009} and references therein). The growth of a bar in the N-body 
simulation with a live halo is a complex process because of the non-linearity in the growing process and its interaction with the dark matter 
halo particles. The growing bars in our simulation are, at times, associated with a nearly constant pattern speed and, at other times, with 
a decreasing pattern speed. Such a growing bar is likely to drive the inner disk to a non-equilibrium state and depending on its strength 
and pattern speed, the inner stellar disk may evolve on $\sim$ a few dynamical time scales. Fundamentally, the growth of the bar facilitates 
the capturing of the disk stars into its various vertical resonances, e.g. the 1:1; 2:1; 4:1 vertical ILRs (inner Lindblad resonances) etc. 
which are often present in the bar region of the disk \citep{Combes1990}. As shown by \citet{Pfenniger1985}, the 4:1 vertical resonance becomes 
stronger as the bar grows, capturing disk stars to form the boxy bulge seen in many N body-simulations, although such studies 
do not explicitly include the interaction of the disk stars with a live dark matter halo. Furthermore, the presence of weakly dissipative orbits in a barred galaxy potential is shown to produce appreciable z-amplification as resonances are crossed, hence facilitating the growth of boxy bulges \citep{Pfenniger1990}. For example, a simple Hamiltonian model of the growth of a bar perturbation \citep{Quillen2002} illustrated the 
effect of resonant capture of the stars into the 2:1 vertical ILR, lifting them out of the disk plane to explain the growth of the boxy/peanut 
bulge. One common result amongst these studies is that the vertical resonances (when present) are efficient in capturing disk stars, facilitating their motion perpendicular to the disk plane, essential for the growth of a boxy bulge. We note that, in this context, the disk stars in our simulation are not only vertically heated at the resonance locations but throughout the bar region of the disk and often with similar magnitude. 
This could imply either the resonant heating is not the only mechanism responsible for the overall vertical heating 
or broad resonances reflecting strong chaos are present throughout the bar region \citep{Pfenniger1985}. Although a growing bar with/without a decreasing pattern speed could promote resonance sweeping through the bar region, it appears unlikely for the disk to accommodate the later scenario making it unclear how the non-resonant stars are vertically heated. In any case, it would be useful to have further insight into how the non-resonant stars might be contributing to the growth of the bulge. 

\subsection{Superthin galaxies}
The disks of many galaxies in our sample are superthin and are dark matter dominated. From the galaxy formation point of 
view, it is extremely important to understand how these superthin galaxies are formed and how they evolve. Can they preserve 
their initial superthinness? In the hierarchical structure formation scenario, a galaxy grows via a large number of mergers 
and as a result, the N-body simulations of galaxy formation underestimate the number of thin galaxies. So it remains a 
puzzle how superthin galaxies remain superthin. Assuming hydrostatic equilibrium, it can be shown that the disk half-thickness 
$h_z \propto \sigma_z/(G \rho_{mid})$; where $\rho_{mid}$ is the mid-plane volume density. Because of various unavoidable 
heating processes $\sigma_z$ would always increase with time and hence the disk thickness ($h_z$). Thus, either superthin 
galaxies somehow maintained their initial thickness or they evolved from an even thinner state to the present day superthin 
state. In the absence of environmental influences, the evolution of these galaxies would depend on the non-axisymmetric 
structures produced in the disk through internal instabilities. Our simulation shows that the non-axisymmetric structures 
such as bars or spirals are spontaneously formed in the disk and heat the disk stars in their respective regime of dominance. 
We briefly mention here that normally in radially hot and vertically very cold galaxies (e.g. mk25, mk27, mk48 for which 
initial $h_z/R_d \sim 0.01 - 0.02$), the disk thickness remains in the superthin regime with the final half thickness 
$h_z/R_d \sim 0.06$ after $5$ to $6$ Gyr of evolution. Most of these galaxies form a {\it type-II} bar but there are exceptions. If 
the initial thickness is greater than $0.03$ or $0.04$, then the final thickness evolves to be in the 'thin' ($h_z/R_d > 0.1$) 
regime.  

\subsection{Common heating mechanism?}
The vertical heating in the outer region is comparatively more complicated because of the presence of spiral arms and, 
at times, mild warps or corrugations. The presence of these perturbations complicate the 
identification of the heating mechanism since more than one may operate simultaneously.  It is known that a stationary 
spiral pattern does not lead to heating (except at resonance locations) and transient spirals heat more effectively in 
the radial direction. However, our analysis indicates that the transient spiral arms also grow (e.g., model mk107) i.e. 
they are time dependent and, hence, a similar situation may prevail in the outer disk as in the central region. In 
both regions, the disk stars respond to the time-dependent potentials. Thus, a common mechanism where the growing bars and 
spirals interact resonantly with the dark matter halo particles and contribute to the vertical heating in their 
respective regions of dominance may be operating.    

\section{Conclusions}
Our investigation of the 70 N-body models of disk galaxies reveals a positive correlation between the growth of a bar and the vertical heating of the disk stars in the central region. A growing bar seems to contribute significantly to the vertical heating of the disk stars. Overall, the heating exponent $\alpha_{in} \geq 1$ for the various galaxy models studied. The disk stars in the central region are generally heated in the vertical direction by a factor $\sim 3$ to $4$ above their initial values over $5$ to $6$ Gyr. 
 
We find that the transient spirals are always present whenever a growing bar is present at the center. Most of these 
spirals are weak and diffuse, and it is likely due to the fact that the disks in our simulations are relatively radially 
hot. The numerical results show that the amount of vertical heating in the outer region is lower compared to the inner 
region with the disk stars in the outer region generally heated vertically by a factor of $\sim 2$ above their 
initial values over a time period of $\sim 4$ Gyr (if there is a growing spiral present) or more (otherwise). 

From the analysis of our simulations, we find that, in general, in radially hot galaxies the vertical heating in the central 
region dominates over radial heating and in the outer region the relative importance is reversed. In contrast, radial heating 
dominates over the vertical heating throughout the disk for radially cold galaxies. We conclude that heating due to 
non-axisymmetric structures appears to be most promising in the context of disk heating problem in general. 

Our simulation results suggest that there is likely a common physical process through which the disk stars are heated 
vertically, which is active throughout the disk from inner to the outer region. Such a process should be investigated in detail 
by studying the vertical motion of the disk stars in the presence of a time dependent perturbing potential which could arise 
due to a growing bar or spiral arms inside a live dark matter halo.  

\medskip
\noindent {\bf Acknowledgement}

All the simulation and analysis presented in this paper are carried out in 3 clusters namely the ASIAA computer cluster, 
TIARA cluster and the Academia Sinica Grid Computing (ASGC) center. The authors would like to express their thanks for 
the generous computing support from all of them. The authors also thank the referee, Daniel Pfenniger, 
for encouraging and useful comments which improved the presentation of the paper. KS would like to thank 
Yao-Huan Tseng, Sam Tseng, and Jason Shih for helping with the computer resources at various stages. KS would like to thank the 
Alexander von Humboldt foundation for its financial support during which the the paper was written.


\end{document}